\documentclass[a4paper, aps, amsmath, 10pt, nofootinbib]{revtex4-2}
\usepackage{amsmath}
\usepackage{amssymb}
\usepackage[Gray,squaren,thinqspace,thinspace]{SIunits}
\usepackage{graphicx}
\usepackage{slashed}
\usepackage{bbm}

\usepackage{xcolor}
\newcommand{\modi}[1]{#1}

\begin{document}

\title{Path-integral approach to Casimir effect with infinitely thin plates}
\author{David Vercauteren}
\email{VercauterenDavid@DuyTan.edu.vn}
\affiliation{Institute of Research and Development, Duy Tan University, Da Nang 550000, Vietnam}
\affiliation{Faculty of Natural Sciences, Duy Tan University, Da Nang 550000, Vietnam}
\begin{abstract}
	When studying the Casimir effect in a quantum field theory setting, one can impose the boundary conditions by adding appropriate Dirac-$\delta$ functions to the path integral. In this paper, the limits of this approach are explored under different boundary conditions.
\end{abstract}

\maketitle

\section{Introduction}
There has been recent interest in the description of the Casimir effect in terms of a quantum field theory on the boundary plates \cite{Dudal:2020yah,Dudal:2024fvo,Dudal:2025qqr,Canfora:2022xcx,Dudal:2024xdu,Canfora:2024awy}. One of the stated goals of this program is the dual superconductor picture of the QCD vacuum \cite{Dudal:2020yah}. This builds on an old idea that the Casimir effect would play an important role in the MIT bag model for hadrons \cite{Milton:1980ke}.\footnote{See chapter 6 of \cite{alma991002315369705502} and references therein for a more recent account.} In this approach, the bag is modeled as a small region of space wherein asymptotic freedom reigns, and with a perfect magnetic conductor as a boundary.

In the usual setup for the Casimir effect, one considers two parallel plates of perfect electrically conducting material. Quantum fluctuation in the vacuum then cause an attractive force to exist between the two plates. The literature abounds \modi{with}\ alternative settings with different shapes of plates, different fields, etc. For the MIT bag model, one considers magnetic conductors instead of electric conductors. In vacuum, electric and magnetic effects are interchangeable due to duality, and so the Casimir effect is not impacted by this modification.

To write down a quantum field theory on the boundary plates, the authors of \cite{Dudal:2020yah,Dudal:2024fvo,Dudal:2025qqr,Canfora:2022xcx,Dudal:2024xdu,Canfora:2024awy} use a formalism developed by Bordag, Robaschik, and Wieczorek in \cite{Bordag:1983zk}. The boundary conditions are imposed by introducing Dirac-$\delta$ functions into the path integral. For ease of reference, I will call this the BRW approach. This approach was also successfully applied in \cite{likardar1,likardar2,Golestanian:1998bx}. \modi{Note}\ that these works have electric conductors as their boundaries.

When applying the BRW approach to computations of the Casimir effect with magnetic conductors for boundaries, or with perfect electromagnetic conductors which interpolate between electric and magnetic, certain divergences appear. The aim of this note is to take a critical look at how those divergences are dealt with in \cite{Dudal:2020yah,Dudal:2024fvo,Dudal:2025qqr,Canfora:2022xcx,Dudal:2024xdu,Canfora:2024awy}. It turns out that the BRW approach must be modified to correctly regularize the infinities whenever the boundaries are not purely electric conductors.

To reduce notational complexity, I will consider massless scalars rather than gauge fields in the following. The motivated reader can easily verify that this only modifies some factors of two from the two physical polarizations of the gauge boson. To make the link between photons and scalars, let us look at what happens to an electromagnetic wave hitting a plate perpendicularly (so the fields parallel to the plate are the ones that matter). In vacuum, it is sufficient to look at the electric field, as the magnetic field can be reconstructed using Maxwell's equations without the matter terms. Assume a plate perpendicular to the $z$-axis.

For a perfect electric conductor, the boundary condition for the electric field is $E_x=0$ and $E_y=0$ on the surface of the plate (in addition to $B_z=0$ for the magnetic field). This means we have Dirichlet boundary conditions for the electric field. For a perfect magnetic conductor, we have $E_z=0$ for the electric field and $B_x=0$ and $B_y=0$ for the magnetic field instead. This implies, through Faraday's law, that $(\vec\nabla\times\vec E)_x = 0$ and $(\vec\nabla\times\vec E)_y = 0$, and thus that $\frac\partial{\partial z}E_x=0$ and $\frac\partial{\partial z}E_y=0$. This means that a magnetic conductor translates to Neumann boundary conditions on the electric field.

Due to duality, we could have made the same argument considering the magnetic field to reach the opposite conclusion. However, it turns out that the BRW approach does not respect duality, as already pointed out in \cite{Karabali:2025olx}. It is possible to redo the following section for a gauge field, and one finds that the computations for electric conductors are identical to those for Dirichlet boundary conditions on the scalar, and similarly magnetic conductors correspond to Neumann boundary conditions. \modi{In section \ref{inter}, I show this in more mathematical detail.}

In section \ref{dir}, I start with a quick review of the Casimir effect with Dirichlet boundary conditions: once in the usual approach by computing the modes, and once in the BRW approach. In section \ref{neu}, I do the same for Neumann boundary conditions, and one finds an untreated divergence in the BRW approach. \modi{Section \ref{inter} clarifies the link between gauge fields and the scalars studied in the rest of the paper.}\ Section \ref{reg} covers several ways to regulate \modi{the divergence found in section \ref{neu}}. In section \ref{wrap}, the results are tied together.

The plates will always be located at $z=0$ and $z=L$. I use Lorentz-index notation $v_\mu$ to write four-dimensional vectors, and vector notation $\vec v$ for vectors in the three-dimensional space-time tangential to the plates. I work in Wick-rotated spacetime (four spatial dimensions) to get rid of bothersome factors of $i$.

\section{Dirichlet Casimir effect} \label{dir}
For self-containedness, let us start with a quick review of the usual Casimir effect.

Assume a massless scalar in 4 dimensions, living in a one-dimensional infinitely deep potential well of length $L$. We have the Lagrangian density
\begin{equation}
    \frac12 \partial_\mu \phi \partial^\mu \phi \;,
\end{equation}
and the boundary conditions
\begin{equation}
    \phi|_{z=0,L} = 0 \;.
\end{equation}
We use the mode expansion
\begin{equation} \label{dirmodes}
    \phi = \sqrt{\frac 2L} \int \frac{d^3p}{(2\pi)^3} \sum_{n=1}^\infty \tilde\phi_n(p) e^{i\vec p\cdot\vec x} \sin\frac{\pi nz}L \;,
\end{equation}
which already fulfills the boundary conditions. The Lagrangian is now
\begin{equation}
    \frac12 \int \frac{d^3p}{(2\pi)^3} \sum_{n=1}^\infty \tilde\phi_n(-p) \tilde\phi_n(p) \left(\vec p^2+\left(\frac{\pi n}L\right)^2\right) \;,
\end{equation}
and the vacuum energy is given by half the logarithm of the determinant of the propagator:
\begin{equation}
    \frac12 \int \frac{d^3p}{(2\pi)^3} \sum_{n=1}^\infty \ln\left(\vec p^2+\left(\frac{\pi n}L\right)^2\right) \;.
\end{equation}
A lot has been written on how to extract a finite value from the above expression. It is generally accepted that dimensional regularization gives the correct result. As it is also (to me at least) the simplest approach, I will use it here.

For future reference, let me say some more on dimensional regularization. The real world has 3+1 dimensions, which we will write as $d$, and we will take $d\to4$ at the end. A plate has 2+1 dimensions, so this will be $d-1$. This means that for $d=3$ (2+1 dimensions), the plates become wires. The Casimir effect for such a setup has been extensively studied, see for example \cite{Ambjorn:1981xw}. \modi{Note}, though, that the number of dimensions perpendicular to the plates is still exactly 1 even in dimensional regularization. We do not take $1-\epsilon$ there.

Let us now proceed with the calculation. The $(d-1)$-dimensional integral is well known, and we find for appropriate values of $d$
\begin{equation}
    - \frac1{2(4\pi)^{(d-1)/2}} \Gamma(-\tfrac{d-1}2) \sum_{n=1}^\infty \left(\frac{\pi n}L\right)^{d-1} = - \frac{\pi^{(d-1)/2}}{2^d} \Gamma(-\tfrac{d-1}2) \frac1{L^{d-1}} \zeta(-d+1) \;.
\end{equation}
This is the regularized vacuum energy, also found by Ambj{\o}rn and Wolfram \cite{Ambjorn:1981xw} (up to a factor of the volume of the plate and with my $d$ equal to their $d+1$). For the Casimir pressure $P_{\text{Cas}}$, take minus the derivative with respect to $L$:
\begin{equation}
    P_{\text{Cas}} = - \frac{\pi^{(d-1)/2}}{2^d} \Gamma(-\tfrac{d-1}2) \frac{d-1}{L^d} \zeta(-d+1) \;.
\end{equation}
This yields finite values for all positive $d$, which is a check on the renormalizability of the setup. For $3+1$ dimensions we get the usual $P_{\text{Cas}} = -\pi^2/480 L^4$.

Now, let us redo the computation using the BRW approach. In this approach, one lets the fields free everywhere in spacetime, and one introduces the boundary conditions through a Dirac-$\delta$ in the path integral:
\begin{equation} \label{pi}
    \int [d\phi] e^{-\tfrac12 \int \partial_\mu \phi \partial^\mu \phi} \delta(\phi(z=0)) \delta(\phi(z=L)) = \int [d\phi] [db^0db^L] e^{-\tfrac12 \int \partial_\mu \phi \partial^\mu \phi + i\int b^0\phi(z=0) + i\int b^L\phi(z=L)} \;,
\end{equation}
where the Fourier representation of the Dirac-$\delta$ function was used. The first integral in the exponential is over all of spacetime, while the second and third integrals are over the plate at $z=0$ and at $z=L$ respectively. Spacetime is infinitely big now, so we use normal, continuous Fourier transforms to rewrite the action as
\begin{equation} \label{dirstart}
    \frac12 \int \frac{d^4p}{(2\pi)^4} p^2 \tilde\phi(-p)\tilde\phi(p) - i \int \frac{d^4p}{(2\pi)^4} \tilde b^0(-\vec p) \tilde\phi(p) - i \int \frac{d^4p}{(2\pi)^4} e^{ip_zL} \tilde b^L(-\vec p) \tilde\phi(p) \;.
\end{equation}
We decouple $\tilde\phi$ from the $\tilde b$s with the shift
\begin{equation}
    \tilde\phi(p) \to \tilde\phi(p) + \frac i{p^2} \left(\tilde b^0(\vec p) + e^{-ip_zL} \tilde b^L(\vec p)\right) \;.
\end{equation}
After this, $\tilde\phi$ can be integrated out of the path integral and gives an $L$-independent determinant that we can ignore. What remains is
\begin{equation} \label{dirrem}
    - \frac12 \int \frac{d^4p}{(2\pi)^4} \left(\tilde b^0(-\vec p)+e^{ip_zL}\tilde b^L(-\vec p)\right) \frac1{p^2} \left(\tilde b^0(\vec p) + e^{-ip_zL} \tilde b^L(\vec p)\right) \;.
\end{equation}
The arguments of the $\tilde b$s do not depend on $p_z$, which can be integrated out. The factor of $1/p^2 = 1/(\vec p^2+p_z^2)$ makes the integral convergent, and we find
\begin{equation}
    - \frac14 \int \frac{d^3p}{(2\pi)^3} \frac1{|\vec p|} \left(\tilde b^0(-\vec p) \tilde b^0(\vec p) + \tilde b^L(-\vec p) \tilde b^L(\vec p) + e^{-L|\vec p|} \tilde b^0(-\vec p) \tilde b^L(\vec p) + e^{-L|\vec p|} \tilde b^L(-\vec p) \tilde b^0(\vec p) \right) \;.
\end{equation}
The vacuum energy is again half the logarithm of the determinant of the propagator:
\begin{equation} \label{dirfin}
    \frac12 \int \frac{d^3p}{(2\pi)^3} \ln \det \begin{pmatrix} 1 & e^{-L|\vec p|} \\ e^{-L|\vec p|} & 1 \end{pmatrix} = \frac1{4\pi^2} \int_0^\infty dp\ p^2 \ln(1-e^{-2Lp}) = - \frac{\pi^2}{1440} \frac1{L^3} \;.
\end{equation}
The pressure is again $-\pi^2/480L^4$.

\section{Neumann Casimir effect} \label{neu}
Encouraged by the results of the previous section, we would like to know what happens in the case we impose Neumann boundary conditions:
\begin{equation}
    \left.\frac d{dz} \phi\right|_{z=0,L} = 0 \;.
\end{equation}
If we use the mode expansion
\begin{equation} \label{neumodes}
    \phi = \sqrt{\frac 2L} \int \frac{d^3p}{(2\pi)^3} \sum_{n=0}^\infty \tilde\phi_n(p) e^{i\vec p\cdot\vec x} \cos\frac{\pi nz}L \;,
\end{equation}
we can repeat the computations from the previous section, and the we conclude that the boundary conditions do not change the Casimir pressure. This is a well-established fact.\footnote{If one imposes different boundary conditions on the two plates, for example Dirichlet at one side and Neumann on the other side, the result is different.}

Let us look, however, at what happens in the BRW approach. The path integral is now
\begin{equation}
    \int [d\phi] e^{-\tfrac12 \int \partial_\mu \phi \partial^\mu \phi} \delta(\tfrac d{dz}\phi(z=0)) \delta(\tfrac d{dz}\phi(z=L)) = \int [d\phi] [db^0db^L] e^{-\tfrac12 \int \partial_\mu \phi \partial^\mu \phi + i\int b^0\tfrac d{dz}\phi(z=0) + i\int b^L\tfrac d{dz}\phi(z=L)} \;.
\end{equation}
In Fourier language, the action is
\begin{equation}
    \frac12 \int \frac{d^4p}{(2\pi)^4} p^2 \tilde\phi(-p)\tilde\phi(p) + \int \frac{d^4p}{(2\pi)^4} \tilde b^0(-\vec p) p_z \tilde\phi(p) + \int \frac{d^4p}{(2\pi)^4} e^{ip_zL} \tilde b^L(-\vec p) p_z \tilde\phi(p) \;,
\end{equation}
and we decouple $\tilde\phi$ from the $\tilde b$s with
\begin{equation}
    \tilde\phi(p) \to \tilde\phi(p) + \frac{p_z}{p^2} \left(\tilde b^0(\vec p) + e^{-ip_zL} \tilde b^L(\vec p)\right) \;.
\end{equation}
What remains in the action after integrating out $\tilde\phi$ is
\begin{equation} \label{pzsq}
    - \frac12 \int \frac{d^4p}{(2\pi)^4} \left(\tilde b^0(-\vec p)+e^{ip_zL}\tilde b^L(-\vec p)\right) \frac{p_z^2}{p^2} \left(\tilde b^0(\vec p) + e^{-ip_zL} \tilde b^L(\vec p)\right) \;.
\end{equation}
The integral over $p_z$ of the terms without the exponential no longer converges due to the extra factor of $p_z^2$. To make this clearer, we can rewrite the previous expression as
\begin{multline} \label{div}
    - \frac12 \int \frac{d^4p}{(2\pi)^4} \left(\tilde b^0(-\vec p) \tilde b^0(\vec p) + \tilde b^L(-\vec p) \tilde b^L(\vec p) + e^{-ip_zL} \tilde b^0(-\vec p) \tilde b^L(\vec p) + e^{ip_zL} \tilde b^L(-\vec p) \tilde b^0(\vec p) \right) \\
    + \frac12 \int \frac{d^4p}{(2\pi)^4} \frac{\vec p^2}{p^2} \left(\tilde b^0(-\vec p) \tilde b^0(\vec p) + \tilde b^L(-\vec p) \tilde b^L(\vec p) + e^{-ip_zL} \tilde b^0(-\vec p) \tilde b^L(\vec p) + e^{ip_zL} \tilde b^L(-\vec p) \tilde b^0(\vec p) \right) \;.
\end{multline}
The second integral converges. In the first integral, integrating the last two terms over $p_z$ gives $\delta(L)$, which is obviously zero. The problem is in the first two terms of the first integral. The integral over $p_z$ is essentially $\int_{-\infty}^{+\infty} \frac{dp_z}{2\pi}$.

In \cite{Dudal:2020yah,Dudal:2024fvo,Dudal:2025qqr}, integrals without the exponential damping $e^{\pm ip_zL}$ are performed by adding this damping and then taking the limit $L\to0$.\footnote{In these papers, they usually consider the integral of $p_z^2 e^{\pm ip_zL}/p^2$ for strictly positive $L$. The condition of strictly positive $L$ results in the absence of the Dirac-$\delta$ term.} This gives
\begin{equation}
    \int_{-\infty}^{+\infty} \frac{dp_z}{2\pi} \to \lim_{L\to0} \int_{-\infty}^{+\infty} \frac{dp_z}{2\pi} e^{\pm ip_zL} = \lim_{L\to0} \delta(L) = 0 \;.
\end{equation}
This makes the first integral vanish, and the second integral then leads to the same results as in the case of Dirichlet boundary conditions. In \cite{Canfora:2022xcx,Dudal:2024xdu,Canfora:2024awy} it is simply stated that $\delta(0)$ is zero in dimensional regularization. However, as stated previously, the $p_z$-integral is exactly 1-dimensional also in dimensional regularization, not $(1-\epsilon)$-dimensional. It is therefore not clear how a dimensional-regularization argument can be valid here.

It is common wisdom in quantum field theory that all proper renormalization procedures should lead to the same physical results. It is therefore worthwhile to look at other ways to regularize the divergent $p_z$ integrals.

\modi{\section{Intermezzo: Link between scalar and gauge boson in BRW} \label{inter}
The goal of this section is to show the link between gauge fields (at tree-level order) and the free scalar, specifically how perfect electric conductors correspond to Dirichlet boundary conditions and perfect magnetic conductors to Neumann boundary conditions. If one uses the approach of solving for the modes in an infinitely deep potential well with appropriate boundary conditions, this correspondence depends on the picture one uses (electric field, magnetic field, gauge field), but the Casimir effect does not depend on the boundary conditions, provided they are identical at both sides. In this section, I will focus on the BRW approach, which uses the gauge field $A_\mu$.

Consider the Lagrangrian density of a (non-Abelian) gauge boson:
\begin{equation}
	\frac14 F_{\mu\nu} F_{\mu\nu} \;,
\end{equation}
where $F_{\mu\nu} = \partial_\mu A_\nu-\partial_\nu A_\mu$ is the usual field strength tensor. To fix the gauge, let us work in a general linear covariant gauge, as is usually the case in the BRW approach (such as in the original paper \cite{Bordag:1983zk}). We add the term $\frac1{2\alpha} (\partial_\mu A_\mu)^2$ to the Lagrangian density, where $\alpha$ is the gauge parameter. The boundary conditions are $i\epsilon_{\mu\nu\kappa\lambda} n_\nu F_{\kappa\lambda} = 0$ for perfect electric conductors and $n_\nu F_{\mu\nu} = 0$ for perfect magnetic conductors. The vector $n_\mu$ is the normal vector on the surface. In our case this is $+\delta_{\mu z}$ at $z=0$ and $-\delta_{\mu z}$ at $z=L$.

Let us start with perfect magnetic conductors. The boundary conditions can be written as $\pm i\vec\nabla\times\vec A = 0$. Remember that vector notation $\vec v$ indicates the part of the fourvector $v_\mu$ tangential to the plates. %Working as in the previous sections, the action in the BRW approach is
In Fourier space the conditions read $\mp \vec p\times\vec{\tilde A} = 0$. There are no conditions for modes with polarization in the direction of $\vec p$ (the three-dimensional projection of the momentum) or in the $z$ direction, nor in any direction in the span of $\vec p$ and $z$. There is a two-dimensional vector space of polarizations that do contribute nontrivially to the Casimir pressure: the ones tangential to the plates but perpendicular to $\vec p$. That we find two polarizations is as expected. I will write these polarizations with Latin indices as $\tilde A_i$. The conditions for these two polarizations are $|\vec p| \tilde A_i = 0$. Furthermore, given that they are perpendicular to the momentum, the Lagrangian density for the $\tilde A_i$s is just $\frac12 p^2\tilde A_i(-p)\tilde A_i(p)$. This means we can start just like in \eqref{dirstart}, but with the $\tilde b$'s replaced with $|\vec p|\tilde b$. This extra factor of $|\vec p|$ will result in an extra factor of $\vec p^2$ in the propagator of the $\tilde b$ fields, but it does not influence the integration over $p_z$, and it contributes only an irrelevant $L$-independent constant to the vacuum energy. One concludes that a perfect electric conductor in the BRW approach is, up to this ultimately irrelevant factor of $\vec p^2$, formally identical to two free scalars with Dirichlet boundary conditions.

For a perfect magnetic conductor, the boundary conditions are $\pm(\vec\nabla A_z-\partial_z\vec A)=0$, or in Fourier language $\pm i(\vec p\tilde A_z-p_z\vec{\tilde A})=0$. Polarizations in the direction of the four-momentum $p_\mu$ are not subject to the boundary conditions, as expected. Polarizations perpendicular to both $p_\mu$ and the $z$-axis have the boundary conditions $\mp ip_z\vec{\tilde A}=0$, or after undoing the Fourier transform $\mp \partial_z\vec A = 0$. This means that those two polarizations obey Neumann boundary conditions. This shows how a perfect magnetic conductor in the BRW approach contains two free scalars with Neumann boundary conditions.

In the case of a perfect magnetic conductor, however, there is a third polarization that contributes nontrivially: the one in the plane spanned by $p_\mu$ and the $z$-axis but perpendicular to $p_\mu$. The existence of this third polarization was also pointed out in \cite{Karabali:2025olx}. To figure out the boundary conditions for this third polarization, we multiply the original boundary conditions for perfect magnetic conductors with $\vec p/|\vec p|$ and use that the polarization of interest is perpendicular to $p_\mu$: $\vec p\cdot\vec{\tilde A} = -p_z\tilde A_z$. This brings us to the boundary conditions $\pm i\frac{p^2}{|\vec p|} A_z = 0$. Undoing the Fourier transform is not practical here. Instead we can dive into the BRW method and start from \eqref{dirstart}, but now with the $\tilde b$'s replaced with $\pm i\frac{p^2}{|\vec p|}\tilde b$. This will result in an extra factor of $-(p^2)^2/\vec p^2$ in the propagator of the $\tilde b$ fields, and thus the factor of $1/p^2$ in \eqref{dirrem} is replaced with $-p^2/\vec p^2$. This finally yields an expression like the first term in \eqref{div}, but with the integrand multiplied with $-p^2/\vec p^2 = -1 - p_z^2/\vec p^2$. As such, the divergence is only worse.

The final lesson is that a perfect magnetic conductor in the BRW approach is identical to two free scalars with Neumann boundary conditions, plus one polarization that diverges even worse.}

\section{Neumann with different regulators} \label{reg}
In this section, we will have another look at the computation of the Casimir effect with Neumann boundary conditions in the BRW approach, and then also in a related approach. The aim is always to ensure that the integrals in \eqref{div} are finite.

There are many regulators we can try. We do need to be careful about the physical implications, as those can sometimes spoil the validity of a regulator. For example, simple cut-off regularization can be problematic for gauge theories as it breaks the gauge invariance, and dimensional regularization is not compatible with supersymmetry. The current computations are for a free scalar, which gives us more leeway. Eventually, though, we are interested in QED and Yang--Mills theory, and so the implications for gauge theory must always be kept in mind.

\subsection{Cut-off}
Cut-off regularization seems to be the simplest option for the problem at hand. For a massless scalar, this should pose no problems. Could we impose a cut-off when looking at a gauge theory?

\modi{Note}\ that the cut-off would be imposed on an integral that relates to the plate dynamics, and not directly to the actual content of the theory. To reword it in more physical language, photons of sufficiently high energy (such as gamma rays) penetrate any material, including conductors. In the BRW approach, the plates are infinitely thin, and so it is natural to assume they would become invisible to photons of sufficiently high energy.\footnote{\modi{It is possible to introduce infinitely thin but still impermeable plates, as in \cite{Saharian:2025ioy}. In that paper, the authors assume that the regions between and outside the plates decouple, and they solve for the modes separately in the different regions. This is different from the BRW approach, where the path integral over the physical fields (gauge fields or the scalar $\phi$) is over functions where $z$ goes from minus to plus infinity.}}

In Yang--Mills theory, a similar argument can be used: the ``plates'' are now a thin slice of the true Yang--Mills vacuum between two MIT bags (hadrons, for example). High-energy gluons can obviously cross such a thin slice, as they are asymptotically free.

A physically correct cut-off would obviously be rather complicated, as the penetration of the high-energy gauge bosons would not be a step function. In practice, though, the details of the regulator should drop out of the final results. Let us then proceed.

Performing the $p_z$-integrals in \eqref{div} with a cut-off $\Lambda$ gives for large $\Lambda$
\begin{multline}
    \frac14 \int \frac{d^3p}{(2\pi)^3} |\vec p| \Bigg(\tilde b^0(-\vec p) \left(1-\frac{2\Lambda}{\pi|\vec p|}\right) \tilde b^0(\vec p) + \tilde b^L(-\vec p) \left(1-\frac{2\Lambda}{\pi|\vec p|}\right) \tilde b^L(\vec p) \\ + e^{-L|\vec p|} \tilde b^0(-\vec p) \left(1-\frac{2e^{L|\vec p|}\sin(L\Lambda)}{\pi L|\vec p|}\right) \tilde b^L(\vec p) + e^{-L|\vec p|} \tilde b^L(-\vec p) \left(1-\frac{2e^{L|\vec p|}\sin(L\Lambda)}{\pi L|\vec p|}\right) \tilde b^0(\vec p) \Bigg) + \mathcal O(\Lambda^{-1}) \;.
\end{multline}
The vacuum energy can be computed the usual way:
\begin{equation}
    \frac12 \int \frac{d^3p}{(2\pi)^3} \ln \det \begin{pmatrix} 1-\frac{2\Lambda}{\pi|\vec p|} & e^{-L|\vec p|}-\frac{2\sin(L\Lambda)}{\pi L|\vec p|} \\ e^{-L|\vec p|}-\frac{2\sin(L\Lambda)}{\pi L|\vec p|} & 1-\frac{2\Lambda}{\pi|\vec p|} \end{pmatrix} = \frac12 \int \frac{d^3p}{(2\pi)^3} \left(C + \mathcal O(\Lambda^{-2})\right) \;.
\end{equation}
Here, $C$ is an $L$-independent expression. The lowest-order terms that depend on $L$ are of order $\Lambda^{-2}$, and so the Casimir pressure vanishes when the cut-off is removed. The best we can say is that the Casimir effect is a UV effect for Neumann boundary conditions in the BRW approach.

\subsection{Thick plates with $b$ fields}
The problem is the infinite extent of the $p_z$-integral in \eqref{div}. We all know that large extent in the momentum domain corresponds to small extent in space, so the problem is actually the infinite thinness of the pla\modi{t}e. And so we are led to ask whether the problem could not be solved with thicker plates. This means that, in the path integral \eqref{pi}, we replace $\delta(\phi(z=0))$ with $\prod_{z=-d}^0 \delta(\phi(z))$ and $\delta(\phi(z=L))$ with $\prod_{z=L}^{L+d} \delta(\phi(z))$, where the products are continuous. Neumann boundary conditions are imposed analogously. This gives both plates a thickness $d$.

It will turn out that both cases, Dirichlet and Neumann, require further regularization. One way to do this is by considering more general Robin boundary conditions instead:
\begin{equation} \label{robincond}
    \left.c\phi - \frac d{dz}\phi\right|_{z=0} = 0 \;, \qquad \left.c\phi + \frac d{dz}\phi\right|_{z=L} = 0 \;, 
\end{equation}
where $c>0$ is a parameter (of dimension mass) that interpolates between Dirichlet and Neumann: $c\to\infty$ gives Dirichlet and $c\to0$ gives Neumann.\footnote{In the case of gauge theory, the analogon of this parameter is dimensionless, and in \cite{Rode:2017yqy} it is called the duality angle, as it interpolates between electric and magnetic conductors.} \modi{T}he normal vector on the plates has different sense left and right, whence the difference in sign.

\modi{For a string vibrating between its two endpoints, Robin boundary conditions can be interpreted as the endpoints being attached to springs perpendicular to the string. If these springs are infinitely strong (spring constant to infinity), this corresponds to Dirichlet boundary conditions ($c\to\infty$), while infinitely weak springs (zero spring constant) correspond to Neumann boundary conditions ($c=0$).

The intuitive interpretation of general Robin boundary conditions in the BRW approach is more muddled. For example, if one approaches the plates from outside ($z<-d$ or $z>L+d$), one finds that the normal vector to the plate is flipped compared to the side between the plates, and the interpretation of \eqref{robincond} is that now $c$ is negative. This would correspond to springs with negative spring constants. The interpretation of \eqref{robincond} inside the plates ($-d<z<0$ and $L<z<L+d$) is even less clear. Only the limit $c\to\infty$ has clear physical meaning throughout: the field $\phi$ is fixed inside the plates. Even Neumann boundary conditions, the other case of interest, cannot readily be understood in terms of strings and springs. This section is therefore best understood as simply a more intuitive renormalization of the BRW approach.}

We can proceed as in the case of infinitely thin plates, and we find for the action after integrating out $\tilde\phi$:
\begin{multline} \label{damp}
	- \frac12 \int\frac{d^4p}{(2\pi)^4} \left(\int_{-d}^0 dz_1 (c-ip_z) e^{ip_zz_1} \tilde b(-\vec p,z_1) + \int_L^{L+d} dz_1 (c+ip_z) e^{ip_zz_1} \tilde b(-\vec p,z_1)\right) \frac1{p^2} \\
    \times \left(\int_{-d}^0 dz_2 (c+ip_z) e^{-ip_zz_2} \tilde b(\vec p,z_2) + \int_L^{L+d} dz_2 (c-ip_z) e^{-ip_zz_2} \tilde b(\vec p,z_2)\right) \;.
\end{multline}
Working out the products once more gives terms with $p_z^2$ like in \eqref{pzsq}, but the thickness of the plates now provides a damping factor $e^{ip_z(z_1-z_2)}$. What were problematic terms in \eqref{div} now give a $\delta(z_1-z_2)$ instead. This can be integrated over, say, $z_2$ to find a finite value. This validates our assumption that a finite thickness would regulate the problematic $p_z$-integral. Doing the integral over $p_z$ gives
\begin{multline}
	- \frac12 \int \frac{d^3p}{(2\pi)^3} \left(\int_{-d}^0 dz + \int_L^{L+d} dz\right) \tilde b(-\vec p,z) \tilde b(\vec p,z) \\
    - \frac14 \int\frac{d^3p}{(2\pi)^3} \frac{c^2-\vec p^2}{|\vec p|} \left(\int_{-d}^0 dz_1 \int_{-d}^0 dz_2 + \int_L^{L+d} dz_1 \int_L^{L+d} dz_2\right) e^{-|z_1-z_2| |\vec p|} \tilde b(-\vec p,z_1) \tilde b(\vec p,z_2) \\
    - \frac14 \int\frac{d^3p}{(2\pi)^3} \frac{(c-|\vec p|)^2}{|\vec p|} \left(\int_{-d}^0 dz_1 \int_L^{L+d} dz_2 + \int_L^{L+d} dz_1 \int_{-d}^0 dz_2\right) e^{-|z_1-z_2| |\vec p|} \tilde b(-\vec p,z_1) \tilde b(\vec p,z_2) \;.
\end{multline}
The first term comes from the $\delta(z_1-z_2)$.

To deal with the $z$-dependence, define
\begin{subequations} \begin{gather}
    \tilde b^\pm(\vec p,z) = \frac1{\sqrt2} \left(\tilde b(\vec p,z+L) \pm \tilde b(\vec p,-z)\right) \qquad (0<z<d) \;, \\
	\tilde b^\pm(\vec p,z) = \sum_{n=-\infty}^{+\infty} e^{2\pi inz/d} \beta_n^\pm(\vec p) \;.
\end{gather} \end{subequations}
The fields $\tilde b^\pm$ are even and odd superpositions of the $b$-fields on both plates. The matrix in \eqref{dirfin} shows that this choice will decouple the two modes. We get for the action in terms of the $\beta$-modes:
\begin{multline} \label{neumannmodes}
	- \frac d2 \int\frac{d^3p}{(2\pi)^3} \sum_\pm \sum_{n=-\infty}^{+\infty} \beta_{-n}^\pm(-\vec p) \beta_n^\pm(\vec p) \\
	- \frac{d^2}4 \int\frac{d^3p}{(2\pi)^3} \frac{c^2-\vec p^2}{|\vec p|} \sum_\pm \Bigg( 2|\vec p|d \sum_{n=-\infty}^{+\infty} \frac1{\vec p^2d^2+4\pi^2n^2} \beta_{-n}^\pm(-\vec p) \beta_n^\pm(\vec p) \\
	+ 2 (e^{-|\vec p|d}-1) \sum_{m,n=-\infty}^{+\infty} \frac{\vec p^2d^2 + 4\pi^2mn}{(\vec p^2d^2 + 4\pi^2m^2)(\vec p^2d^2 + 4\pi^2n^2)} \beta_m^\pm(-\vec p) \beta_n^\pm(\vec p) \Bigg) \\
	- \frac{d^2}4 \int\frac{d^3p}{(2\pi)^3} \frac{(c-|\vec p|)^2}{|\vec p|} e^{-L|\vec p|} (e^{-|\vec p|d}-1)^2 \sum_\pm (\pm1) \sum_{m,n=-\infty}^{+\infty} \frac{\beta_m^\pm(-\vec p)}{|\vec p|d-2\pi im} \frac{\beta_n^\pm(\vec p)}{|\vec p|d-2\pi in} \;.
\end{multline}
Now define the following matrices and vectors:
\begin{subequations} \begin{gather}
	M_{mn} = \left(\frac{2|\vec p|d}{\vec p^2d^2+4\pi^2m^2}+\frac{2|\vec p|}{d(c^2-\vec p^2)}\right) \delta_{m+n} \;, \\
	a_m = \sqrt{2 (e^{-|\vec p|d}-1)} \frac{|\vec p|d}{\vec p^2d^2 + 4\pi^2m^2} \;, \\
	b_m = \sqrt{2 (e^{-|\vec p|d}-1)} \frac{2\pi m}{\vec p^2d^2 + 4\pi^2m^2} \;, \\
	c_m^\pm = s^\pm \sqrt{\frac{c-|\vec p|}{c+|\vec p|}} e^{-\frac L2|\vec p|} (e^{-|\vec p|d}-1) \frac1{|\vec p|d-2\pi im} \;,
\end{gather} \end{subequations}
where $s^+=1$ and $s^-=i$.\footnote{The signs of the vectors $a$, $b$, and $c$ can be chosen arbitrarily; the final result does not depend on such choices. The choice of branch cut for the square roots is irrelevant by the same reason.} This allows us to rewrite the action as
\begin{equation}
	- \frac{d^2}4 \int\frac{d^3p}{(2\pi)^3} \frac{c^2-\vec p^2}{|\vec p|} \sum_\pm \sum_{m,n=-\infty}^{+\infty} \beta_m^\pm(-\vec p) \left(M_{mn} + a_m a_n + b_m b_n + c_m^\pm c_n^\pm\right) \beta_n^\pm(\vec p) \;.
\end{equation}
We now need to compute the logarithm of the determinant of $M_{mn} + a_m a_n + b_m b_n + c_m^\pm c_n^\pm$. The matrix $M$ is diagonal and $L$-independent, and so we can use that $\ln\det(M_{mn} + a_m a_n + b_m b_n + c_m^\pm c_n^\pm) = \ln\det M + \ln\det(\delta_{mn} + M^{-1}_{mk}a_k a_n + M^{-1}_{mk}b_k b_n + M^{-1}_{mk}c_k^\pm c_n^\pm)$. The first term is an $L$-independent constant that we can ignore. For the second term, \modi{it is easy to see}\ that any vector perpendicular to $a$, $b$, and $c$ is a right eigenvector of eigenvalue 1, and any vector perpendicular to $M^{-1}a$, $M^{-1}b$, and $M^{-1}c$ is a left eigenvector of eigenvalue 1. The eigenvalues of interest come from the spans of $a$, $b$, and $c$ (on the right) and of $M^{-1}a$, $M^{-1}b$, and $M^{-1}c$ (on the left). It is tedious but straightforward to compute the operator in question in the relevant vector spaces, and the result is that
\begin{equation}
	\ln\det(M_{mn} + a_m a_n + b_m b_n + c_m^\pm c_n^\pm) = \ln\det M + \ln\det\begin{pmatrix} 1+aM^{-1}a & aM^{-1}b & aM^{-1}c^\pm \\ bM^{-1}a & 1+bM^{-1}b & bM^{-1}c^\pm \\ c^\pm M^{-1}a & c^\pm M^{-1}b & 1+c^\pm M^{-1}c^\pm \end{pmatrix} .
\end{equation}
Computing all the matrix and vector products and simplifying yields for the vacuum energy up to $L$-independent terms\footnote{Some of those $L$-independent terms are singular in both the pure-Dirichlet limit $c\to\infty$ and the pure-Neumann limit $c\to0$, which shows how introducing $c$ was necessary to regulate the intermediate computations.}
\begin{equation}
    \frac12 \sum_\pm \int \frac{d^3p}{(2\pi)^3} \ln\left(1\pm e^{-L|\vec p|} \frac{1-e^{-2cd}}{\left(\frac{c+|\vec p|}{c-|\vec p|}\right)^2-e^{-2cd}}\right) \;.
\end{equation}
In the pure-Dirichlet limit $c\to\infty$, this reduces to \eqref{dirfin} as it should. This means that all $d$-dependence drops out in this limit, and the BRW approach with infinitely thin plates from the start is justified for Dirichlet boundary conditions.

If instead we take the pure-Neumann limit $c\to0$, and we expand for small $d$, the Casimir pressure evaluates to $-\frac{15d^2}{64\pi^2L^6} \left(1-\frac{3d}L+\cdots\right)$. For general Robin boundary conditions, we find for small $d$ that the Casimir pressure is equal to $-\frac{15d^2}{64\pi^2L^6} \left(1-\frac85 c L + \frac65 c^2 L^2 - \frac8{15} c^3 L^3 + \frac2{15} c^4 L^4\right)+\cdots$. \modi{T}he limits $d\to0$ and $c\to\infty$ do not commute.

As such, we once more find that the Casimir pressure for Neumann boundary conditions \modi{in the BRW approach}\ vanishes for very thin plates. \modi{I}n the limit $d\to\infty$, \modi{however,}\ one does not recover the expected result (see \cite{deAlbuquerque:2003qbk} for two plates with identical Robin boundary conditions, called RR there, and with their $a$ equal to my $L$). If one assumes $c<0$, the limit $d\to\infty$ recovers \eqref{dirfin}, which would be the expected result if one changes the sign of $c$ at one of the two plates in \eqref{robincond}, as is easily verified adapting the mode expansion \eqref{dirmodes} and continuing the computations there for such a case. If one redoes the computations in this subsection but with such a changed sign, one amusingly does recover the results of \cite{deAlbuquerque:2003qbk}. It seems that the BRW approach with thick plates and for Robin boundary conditions is confused about the sign of $c$. \modi{This may be related to the fact that the boundary conditions have the wrong sign outside the plates, as mentioned in the beginning of this section.}

\subsection{Static, non-dynamical background}
In \cite{Graham:2003ib}, Graham \emph{et al.} propose a different approach. Instead of imposing the boundary conditions with an addition to the path integral, they add terms to the action. They consider Dirichlet boundary conditions and impose them with an extra term
\begin{equation}
    \frac12 \int d^4x\ \sigma(x) \phi(x)^2 \;.
\end{equation}
Here, $\sigma$ is a nondynamic background field. If one chooses $\sigma$ to be a sum of rectangular/boxcar functions, then the boundary conditions are imposed in the limit of infinitely high rectangles. Concretely, the pair of thick plates considered in the previous subsection would be parametrized by
\begin{equation} \label{sigmadef}
    \sigma(x) = \lambda (H(z+d)H(-z) + H(z-L)H(-z+L+d)) \;,
\end{equation}
where $H$ is the Heaviside step function. In the limit $\lambda\to\infty$, such a term forces $\phi(x)^2$ to be zero inside the plates.\footnote{In \cite{Milton:2004ya} a similar approach is used, but with $\sigma(x) = \frac\lambda L(\delta(z)+\delta(z-L))$. It seems that this simplifies the computations of the vacuum energy, but the approach by Graham \emph{et al.} allows for the more intuitive explanation of what is going on in the next section.}

If such a term is added to the action, one needs to reconsider the renormalization of the theory. In the Dirichlet case, new divergences are generated of the form $\sigma$ and $\sigma^2$, see \cite{Graham:2003ib} for a full account. Those divergences are local, though, and so do not depend on $L$. As such, they will not influence the Casimir pressure.

In order to compute the vacuum energy, one needs the determinant of the operator $-\partial^2 + \sigma$, or after Fourier transforming the dimensions transverse to the plate
\begin{equation} \label{dirgraham}
    - \frac{d^2}{dz^2} + \vec p^2 + \sigma(z) \;.
\end{equation}
Graham \emph{et al.} use phase shifts to compute this, but it is possible to cut out a step using the Gel'fand--Yaglom theorem for one-dimensional functional determinants in a finite interval \cite{Gelfand:1959nq,Dunne:2007rt}. To do so, put the system in a finite box from $z=-Z$ to $z=+Z$. Then one solves the initial value problem
\begin{equation} \label{ivp}
    \left(- \frac{d^2}{dz^2} + \vec p^2 + \sigma(z)\right) \phi(z) = 0 \;, \qquad \phi(-Z) = 0 \;, \qquad \phi'(-Z) = 1 \;,
\end{equation}
and
\begin{equation} \label{GYdet}
    \det\left(- \frac{d^2}{dz^2} + \vec p^2 + \sigma(z)\right) \propto \phi(+Z) \;.
\end{equation}
This is a proportionality, and to get a definite result, one divides by a determinant of a different operator. For the case at hand, we then take the limit $Z\to\infty$. It is obvious that, as a shortcut, we can put $Z\to\infty$ from the start, demand $\phi(z) \sim e^z$ for $z\to-\infty$, and look at the coefficient of the exponential divergence at $z\to+\infty$.

For $\sigma(z)$ of the form \eqref{sigmadef}, we find that the solutions to \eqref{ivp} are linear combinations of $e^{\pm|\vec p|z}$ between and outside the plates, and linear combinations of $e^{\pm z\sqrt{\vec p^2+\lambda}}$ inside the plates. Solving the initial value problem \eqref{ivp} then amounts to gluing those solutions together. This is an exercise in systems of two linear equations in two unknowns (namely the coefficients in the linear combinations). Doing so for all the transitions, one finds $\phi(+Z)$. Taking $Z\to\infty$ and then dividing by the same determinant for $L\to\infty$ (to get rid of the proportionality constant in \eqref{GYdet}) yields the determinant, up to an $L$-independent coefficient. Half the logarithm gives the vacuum energy, again up to irrelevant $L$-independent terms:
\begin{equation}
    \frac12 \int \frac{d^3p}{(2\pi)^3} \ln \left( 1 - \frac{e^{-2|\vec p|L}\lambda^2}{\left(2\vec p^2+\lambda+2|\vec p|\sqrt{\vec p^2+\lambda}\coth d\sqrt{\vec p^2+\lambda}\right)^2} \right) \;.
\end{equation}
In the limit $\lambda\to\infty$, this gives the usual result for Dirichlet boundary conditions.

To modify the approach for Neumann boundary conditions, we need to add the term
\begin{equation}
    \frac12 \int d^4x\ \sigma(x) \left(\frac d{dz}\phi(x)\right)^2
\end{equation}
to the action instead. Such an addition will obviously lead to a great many extra primitive divergences, but in the free theory we are considering here, those can all be absorbed by local counterterms of the form $\sigma^n$. Again, contributions from such counterterms to the vacuum energy do not depend on $L$ due to their locality, and we can proceed without having to actually compute them.

We now need the determinant of the operator
\begin{equation}
    - \frac d{dz} (1+\sigma(z)) \frac d{dz} + \vec p^2 \;.
\end{equation}
This is complicated by the existence of points where $\sigma'(z)\neq0$. However, we can rewrite this as $\det(- \frac d{dz} (1+\sigma(z)) \frac{d^2}{dz^2} + \vec p^2 \frac d{dz}) / \det(\frac d{dz}) = \det(- (1+\sigma(z)) \frac{d^2}{dz^2} + \vec p^2 )$. Now, the determinant of $1+\sigma(z)$ does not depend on $L$ (we have $\ln\det(1+\sigma(z)) = \int dz \ln(1+\sigma(z)) =2d\ln\lambda$). We can therefore equivalently consider the operator
\begin{equation} \label{neugraham}
    - \frac{d^2}{dz^2} + \frac{\vec p^2}{1+\sigma(z)} \;.
\end{equation}
The initial value problem is solved by linear combinations of $e^{\pm|\vec p|z}$ between and outside the plates, and linear combinations of $e^{\pm|\vec p|z/\sqrt{1+\lambda}}$ inside the plates. We proceed as in the Dirichlet case and find for the vacuum energy up to irrelevant $L$-independent terms:
\begin{equation}
    \frac12 \int \frac{d^3p}{(2\pi)^3} \ln \left( 1 - \frac{e^{-2|\vec p|L}\lambda^2}{\left(\lambda+2+2\sqrt{1+\lambda}\coth\frac{|\vec p|d}{\sqrt{1+\lambda}}\right)^2} \right) \;.
\end{equation}
This time, the $\lambda\to\infty$ limit gives
\begin{equation}
    \frac12 \int \frac{d^3p}{(2\pi)^3} \ln \left( 1 - \frac{e^{-2|\vec p|L}}{(1+\frac2{|\vec p|d})^2} \right) \;.
\end{equation}
This once more gives $-\frac{15d^2}{64 \pi^2 L^6} \left(1-\frac{3d}L+\cdots\right)$ for the Casimir pressure at small thickness $d$. Again, the Casimir pressure vanishes for infinitely thin plates \modi{in this approach}. \modi{T}he limit $d\to\infty$ \modi{also yields}\ the expected result \eqref{dirfin}.

This is consistent with what was found in \cite{Fosco:2009ic}. They also impose the boundary conditions by adding terms to the action, initially encoding infinitesimally thin plates. They then also find that a finite plate thickness is required to really impose Neumann boundary conditions in their approach.

\section{Wrap-up} \label{wrap}
From the previous section, it is obvious that putting $\frac d{dz}\phi=0$ in vanishingly thin plates does not yield any Casimir effect\modi{, neither when using the BRW approach nor in the approach using static background fields \emph{à la} Graham \emph{et al}}. \modi{(As mentioned before, t}his conclusion does not \modi{hold}\ for approaches such as in \cite{Saharian:2025ioy}. They do consider thin plates, but they solve for the modes between the plates and outside the plates separately. This means that the modes are implicitly interrupted in the bulk of the plates\modi{, and the system is effectively put into three infinitely deep potential wells: one from minus infinity to the first plate, one between the plates, and one from the second plate to plus infinity.)}\ We went through three different ways to regulate the problem: BRW with cut-off, BRW with thick plates, and the background approach by Graham \emph{et al.} In all three cases, the conclusion is that the Casimir pressure goes to zero for infinitely thin plates\modi{, except in the Dirichlet case}. In the last two cases, we found exactly the same small-thickness behavior, even though the approaches were quite different, and we did recover the usual values for the Casimir pressure when taking the plate thickness to infinity (but not for general Robin boundary conditions). This lends credence to the correctness of the results obtained there.

It appears that the regulators used in \cite{Dudal:2020yah,Dudal:2024fvo,Dudal:2025qqr,Canfora:2022xcx,Dudal:2024xdu,Canfora:2024awy} are faulty\modi{. One cannot compute the integral $\int_{-\infty}^{+\infty}\frac{dp_z}{2\pi} \frac{p_z^2}{p^2}$ by integrating $p_z^2 e^{\pm ip_zL}/p^2$ and taking $L\to0$ as done in \cite{Dudal:2020yah,Dudal:2024fvo,Dudal:2025qqr}. This is because the integral of $p_z^2 e^{\pm ip_zL}/p^2$ is not continuous around $L=0$ and has a term $\propto \delta(L)$ there. And one cannot put $\delta(0)=0$ as done in \cite{Canfora:2022xcx,Dudal:2024xdu,Canfora:2024awy}, as the argument of dimensional regularization is not valid (there is exactly one direction perpendicular to the plate, never $1-\epsilon$). The more physical ways of taming this divergence explored in the previous section show how this divergence in the integral over $p_z$ does influence physical results and leads to expressions incompatible with what is found in the usual approach, namely by putting the system in an infinitely deep potential well with the appropriate boundary conditions.}

To get some intuition for what exactly is happening, let us revisit the approach by Graham \emph{et al.}

Consider a wave traveling along the $z$-axis under the Hamiltonian in \eqref{dirgraham}, and hitting one of the plates. Inside the plate, the wave function will go like $e^{\pm z\sqrt\lambda}$ for very large $\lambda$. The meaning hereof is clear: in the case of a plus sign, we interpret this as a wave coming from the right, being strongly damped by the wall, and coming out at the left side reduced by a factor of $e^{-d\sqrt\lambda}$. The case of a minus sign is the same but in opposite sense. In the limit $\lambda\to\infty$, the plates become completely opaque and no waves can go through. This shows that the Graham \emph{et al.} approach is a good way to impose Dirichlet boundary conditions.

If instead we aim for Neumann boundary conditions, we are led to the Hamiltonian in \eqref{neugraham}. In the limit of large $\lambda$ this is just $-d^2/dz^2$, which is solved by any linear function of $z$. This means that a wave hitting the plate will go through linearly and continue at the other side. It may be reduced in amplitude, but even in the limit $\lambda\to\infty$, most waves will go through in some form or other. If we then take the limit of small $d$, the plates become harmless: all waves pass through without modification. This is why the Graham \emph{et al.} approach as is does not work for Neumann boundary conditions.

Infinitely thick plates solve the problem: A linearly rising function is not square integrable and must be discarded. Therefore, the only waves we can keep are the ones that hit the plates with exactly $\frac d{dz}\phi=0$, and those are the ones considered in \eqref{neumodes}.

If we now turn our attention to QED or Yang--Mills, we find that, for finite plate thickness, there is no electric--magnetic duality any longer. From the considerations just given, the reason is obvious: For finite plate thickness, waves can enter the plates, which must be made of matter and therefore break duality. Duality can only be seen in cases where waves penetrating the plates have no impact on the final physical results.

In a true conductor this would be the case even for finite thickness: In addition to boundary conditions like $\vec E_{\|} = 0$, a true perfect conductor also has bulk conditions like $\vec E=0$ inside the plate. And it appears that, for perfect electromagnetic conductors that are not purely of electric type, this matters.

Nevertheless, the papers \cite{Dudal:2020yah,Dudal:2024fvo,Dudal:2025qqr,Canfora:2022xcx,Dudal:2024xdu,Canfora:2024awy} somehow did obtain the expected results. It is not immediately clear to me why this should happen. But until this can be clarified, results obtained from the BRW approach with Neumann or Robin boundary conditions (and no bulk conditions) plus dropping of infinities should be treated with caution.

\section*{Acknowledgments}
The author would like to thank David Dudal for constructive discussions and for drawing attention to several papers that greatly informed the work here presented.

\bibliographystyle{unsrt}
\bibliography{Bibliografie}

\end{document}